\documentclass[10pt]{article}
\usepackage[english]{babel}
\usepackage{graphicx}

\def\gee{ \, \lower 1mm\hbox{$\,{\buildrel > \over{\scriptstyle\scriptstyle\sim} }\displaystyle \,$}}
\def\lee{ \, \lower 1mm\hbox{$\,{\buildrel < \over{\scriptstyle\scriptstyle\sim} }\displaystyle \,$}}
\def\|{\partial}

\def\varkappa {{\scriptstyle\partial}\! e}

\begin{document}
\begin{center}
 SELF-CONSISTENT GAS AND STELLAR DYNAMICS OF DISK GALAXIES: A
PROBLEM OF DARK MASS
\end{center}

\begin{center}
 Alexander V. Khoperskov, Sergej S. Khrapov
\end{center}

\begin{center}
 \textit{Volgograd State University, Russia, E-mail}: khoperskov@volsu.ru
\end{center}

\begin{abstract}\noindent
 We present\footnote{\normalsize\underline{2006ASSL..337..337K}~: Astrophysics and Space Science Library, 2006, v.337, p.
 337--344 //
 Astrophysical Disks / Alexei M. Fridman, et al (eds.). Publ. by Springer, Dordrecht, The Netherlands, 2006}
  results of numerical modeling made for the galactic stellar and stellar-gas disk embedded in the spherical halo and bulge. The stellar disk is simulated by N-body system, the equations of hydrodynamics are solved by TVD-method. We used TREEcode-algorithm for calculation of a self-gravity in stellar and gaseous components. The possibility of bars birth in a hot stellar disk because of gravitational instability of a cold gas component is investigated. The conditions of occurrence lopsided-galaxies from a axisymmetric disk as a result of gravitational instability are explored. The self-consistent models of double bars are constructed and the dynamical stability of these structures is discussed.
\end{abstract}



\section{Introduction}

Dynamics of many structures in disk galaxies is considerably
determined by spherical subsystem properties and, in particular,
by characteristics of density distribution of dark halo in stellar
disk limits.

 The bar formation because of global bar-mode instability is impeded, if halo mass inside of optical radius surpasses disk mass
 $M_h\lee (1-1.5) \cdot M_d $ [\ref{Bisnovatyi-Kogan-1984}, \ref{Polyachenko-1979}].
 On the other hand, the observations data and N-body simulation give estimates $M_h/M_d > 1.5 $ for some galaxies
[\ref{Khoperskov-etal-2001}, \ref{Khoperskov-2002}].
  Gas component is cold because of radiative cooling and can be gravitationally unstable. The unstable modes in massive gas disk are capable to generate the bar even in the hot stellar disk in case of a high dispersion of stars velocities and at presence of the massive halo.

A asymmetrical spiral structure (one-arm) and bar displacement
concerning disk centre are typical distinctive features of a
series SBcd--SBm galaxies (Magellanic type).
  These properties are observed at LMC, NGC 55, 925, 1313, 1744, 4490, 4618, 4625 etc.
  [\ref{Odewahn-1991}, \ref{Pisano-2000}, \ref{Vaucouleurs-1972}].
  The formation mechanism of the displaced bar and other features of lopsided-galaxies can be caused by preferred growth of one-arms modes in gravitationally unstable disk and by subsequent interaction of these perturbations with a bar-mode at a nonlinear stage
   [\ref{Zasov-2002}].
The late type galaxies contain more gas, than early type objects.
Therefore question on influence of gas on the bar displacement and
asymmetry in disk structure requires of special study.

The small-scale asymmetrical structures at disks center are very
important for understanding of a phenomenon of nuclear galactic
activity. The double bars can deliver gas to a active nuclei
[\ref{Shlosman-1989}].
 The photometric data are the basic evidence about presence of the second inner bar at approximately 70 galaxies
  [\ref{Moiseev-2001}].
 The self-consistent models of double bars were studied by the N-body method
 [\ref{Friedli-1993}, \ref{Khoperskov-Moiseev-2001}].
 Key problem of double bars is the question on dynamic stability of these systems.

\section{Modelling}

\subsection{The numerical model of stellar-gas disk}

\begin{figure}[!t]
{\includegraphics[width=0.49\hsize]{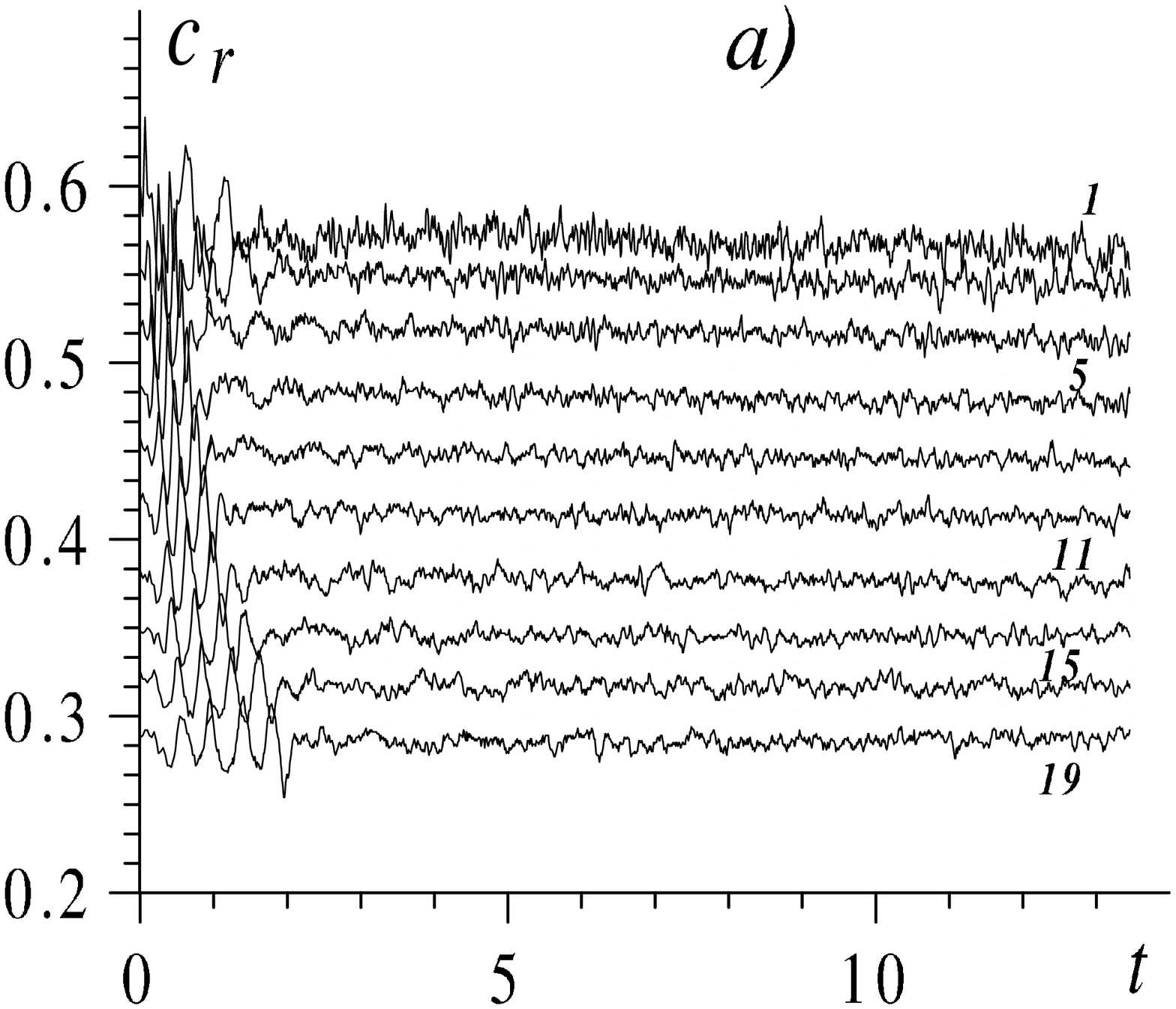}}
 \vskip -0.425\hsize \hskip 0.5\hsize
{\includegraphics[width=0.49\hsize]{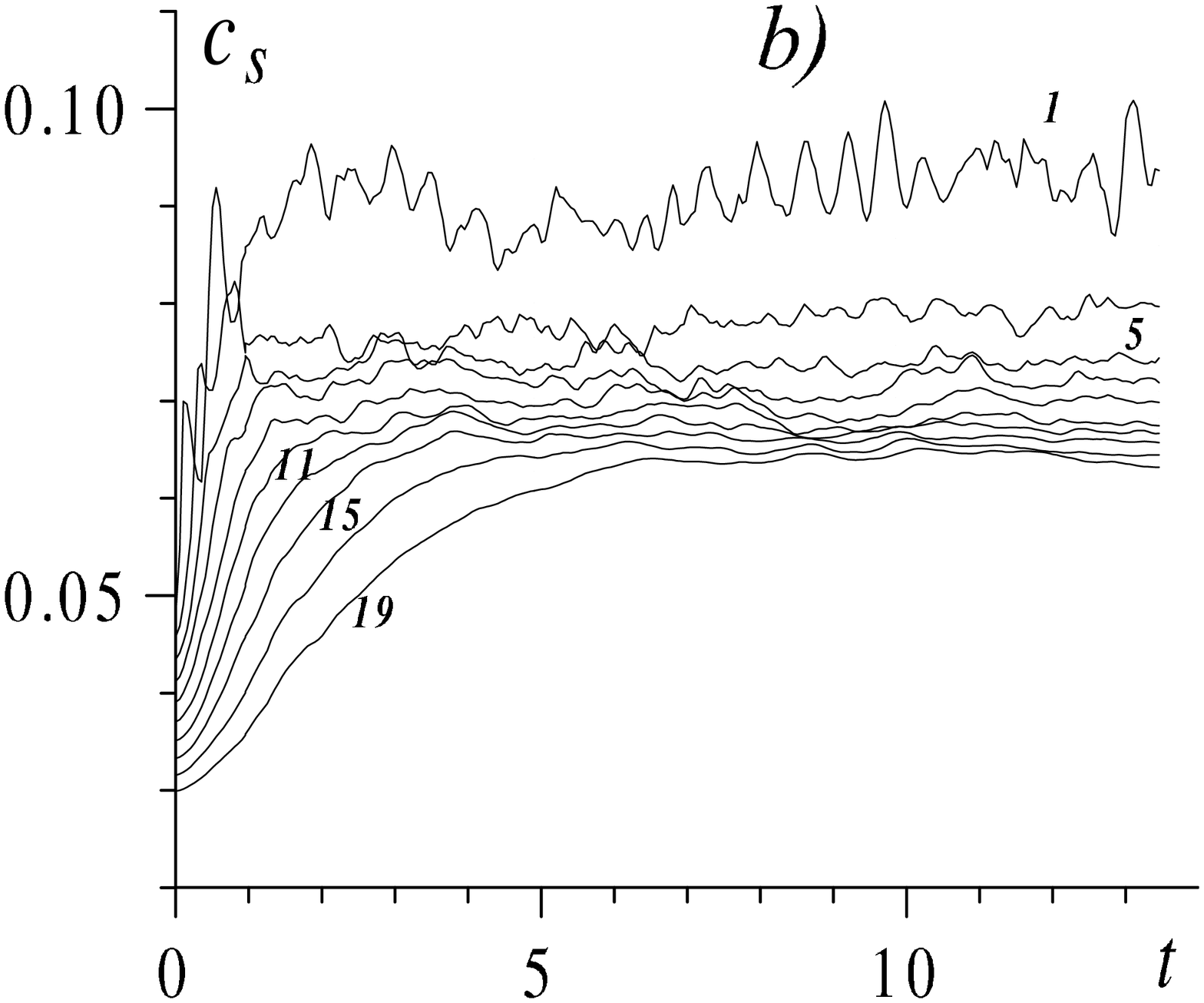}}\vskip -0.02\hsize
  \caption {
Time dependences of the dispersion of stars radial velocities
$c_r$ ({\it a}) and sound speed of gas  $c_s $ ({\it b}) on
various radiuses (curve {\it 1} --- disk center, {\it 2}
--- disk periphery).
 }\label{AVH-SG-Cr-Cs}
\end{figure}

 The 3D stellar disk simulation is based on  $N$-body model, taking into account an external field of the rigid matter distributions in bulge and halo. The gas disk model is constructed on the non-viscous equations of gasdynamics, and is complemented by gravitational forces on the part of stellar disk, spherical subsystem, and gas self-gravity
 also.

 The gas galactic disks are cold, as the sound velocity
 $c_s$ is much less than the dispersion of radial velocities in stellar disks $c_r$.
 The radiative losses in the equation on energy are defined by quantity
 $Q^{-}$:
 \begin{equation}\label{4--Eq-cooling-gas}
Q^{-}=A_{c}
\,\frac{(c_s^2-c_{s1}^2)^\alpha}{(c_{s2}^2-c_{s}^2)^\beta}\cdot
\varrho^2 \  \textrm{for} \ c_s > c_{s1} \ \ \
 (Q^{-}=0 \ \textrm{for} \ c_s < c_{s1}) \,,
\end{equation}
 $ \varrho $ --- density, parameters $A_{c}$, $c_{s1}$,
 $\alpha$, $c_{s2}$, $\beta$ are free.
 The cooling of gas strongly grows in the case $c_s \rightarrow c_{s2}$,
 therefore restriction $c_s < c_{s2}$ is carried out always.
We solved hydrodynamical equations by the method TVD-E. The
self-gravity account in gas and stellar disks is based on
TREEcode. We simulated disks with an exponential profile of
surface density and radial scale $L$.

 If the stellar disk is on threshold of gravitational stability, small mass of cold gas component does not give to an additional heating of the stellar system
 (fig.~\ref{AVH-SG-Cr-Cs}).
The gas mass is equal  $M_g=0.08\cdot M_d$ inside the stellar disk
in this model. And the massive halo ($M_h=3M_d$) forbids the bar
formation, as in stellar component (which besides is hot), and in
gas disk.

\subsection{The stellar bar formation because of gravitational instability in the gas disk}

\begin{figure}[!t]
\centerline{\includegraphics[width=1.01\hsize]{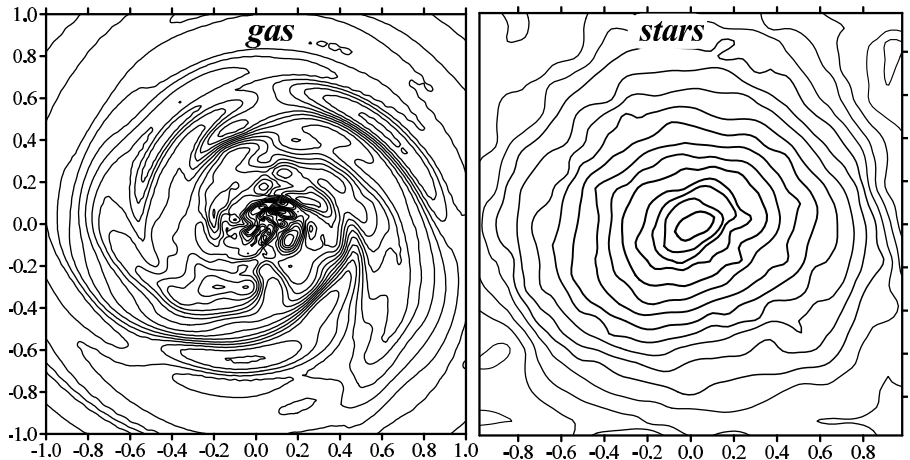}}
 \vskip 0 mm
\caption{
 The gravitational instability of gas is the reason of bar formation in central area of the hot stellar disk. The isolines of a surface density are shown.
}\label{AVH-SG-Bar1}
\end{figure}

Let's consider models with halo mass  $M_h=(1\div 2.5)\cdot M_d$.
 In all cases the initial dispersion of star velocities and the massive halo provide gravitational stability
 of the stellar disk in absence of gas. The account of gas can qualitatively change evolution of
 system.

\begin{figure}[!t]
{\includegraphics[width=0.5\hsize]{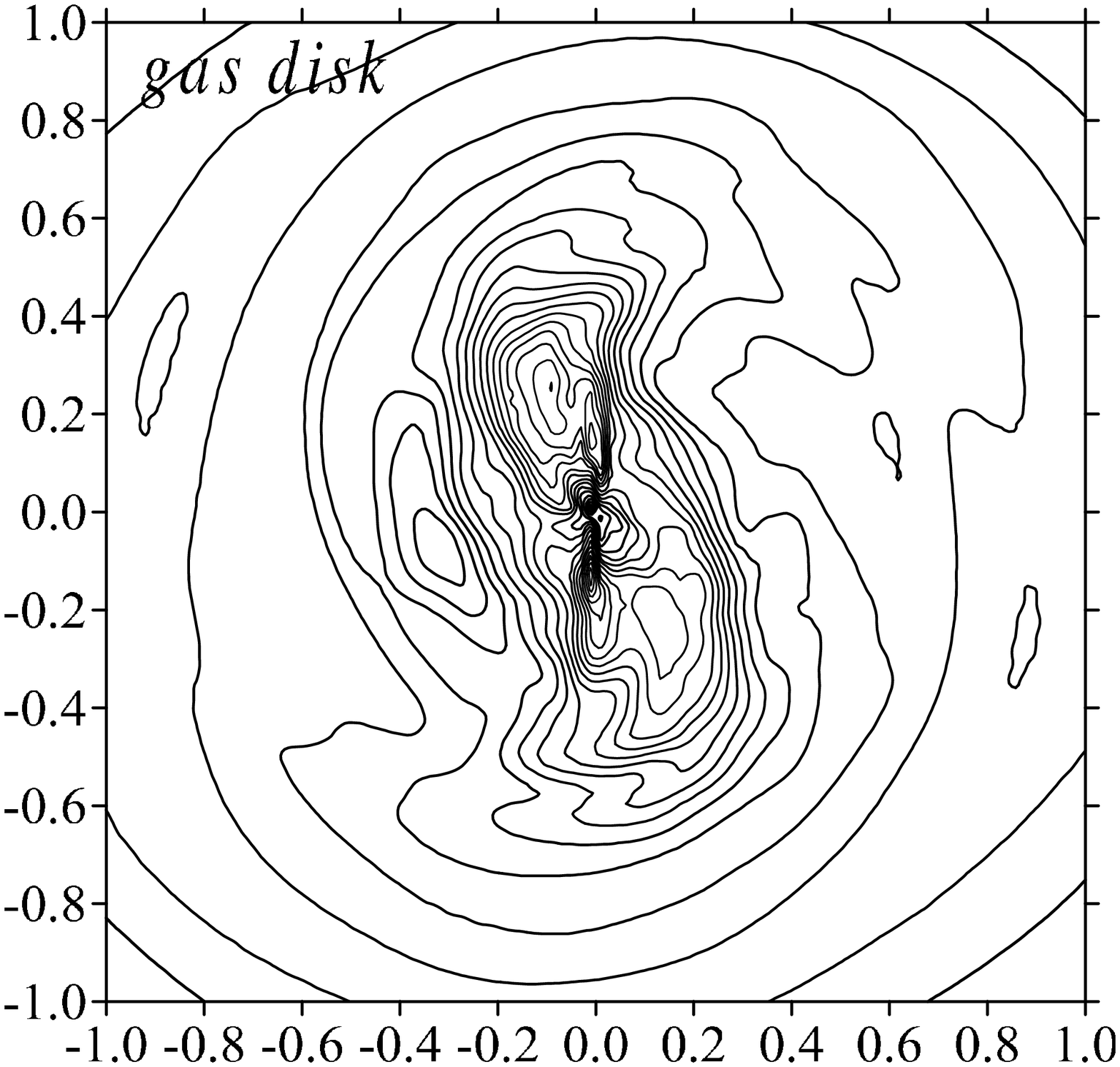}}
 \vskip -0.51\hsize \hskip 0.5\hsize
{\includegraphics[width=0.495\hsize]{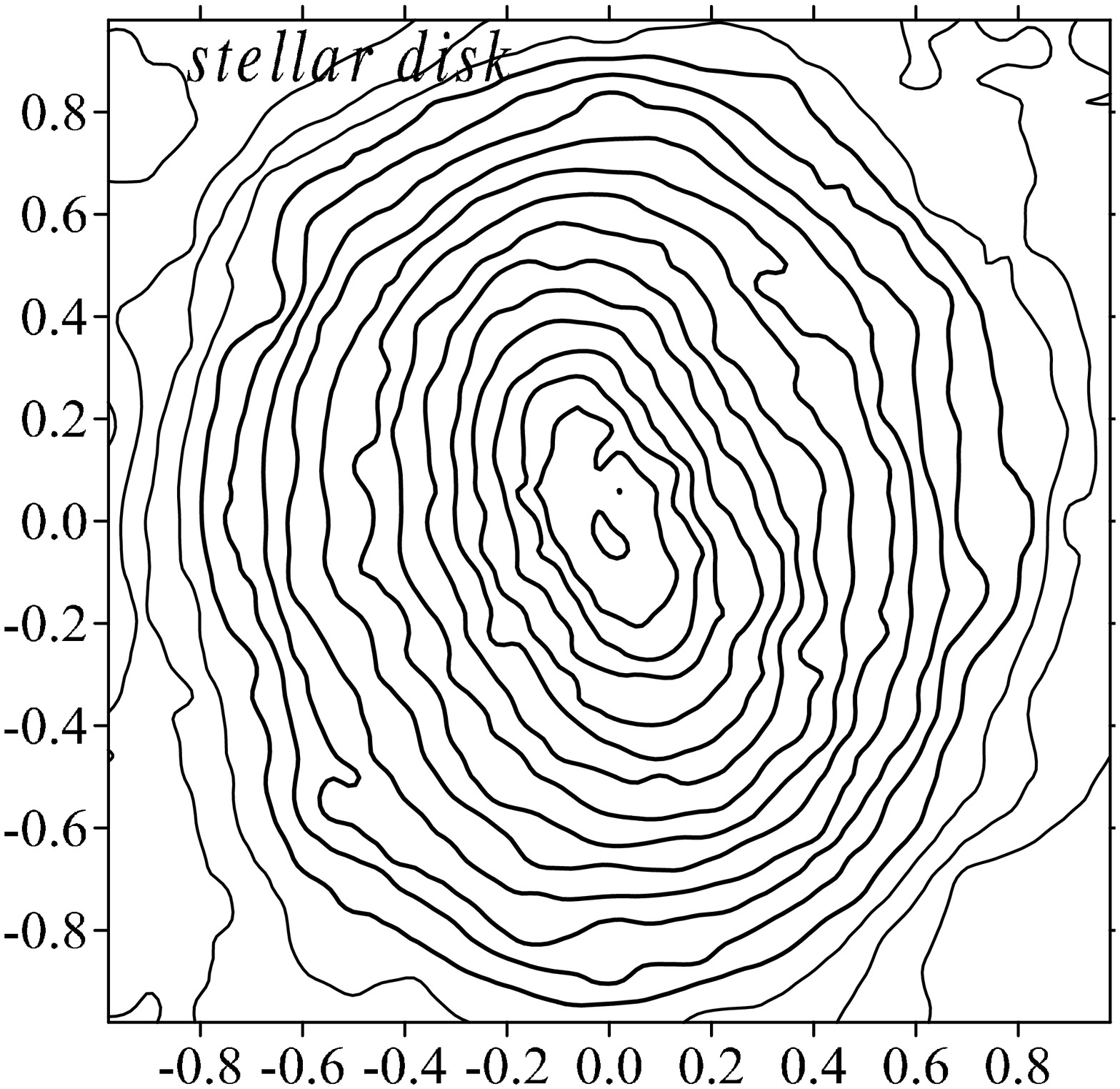}}
 \caption{
 The isolines of surface density of stellar disk (left panel) and gaseous disk (right panel) after 5 rotation periods of outer part of disk. The standings of shock waves are well visible in region of the
 bar.
}\label{AVH-SG-Bar2}
\end{figure}

 The radiative cooling provides cold, gravitationally unstable state of gas component, it gives in formation in gas of non-axisymmetric structures, which in turn generate disturbances in the stellar disk
  (fig.~\ref{AVH-SG-Bar1}).
 There is the prompt bar formation in the stellar disk because of gravitational gas instability, if the models contain a lot of gas. It is important, that the instability of gas can generate the bar in hot stellar disk
  (fig.~\ref{AVH-SG-Bar2}), when Toomre's parameter exceeds $Q_T=c_r/c_T\gee 2 $ in the region
$r\le L$.
 Such disks are stable without gas.

\subsection{Lopsided-galaxies}

Let's consider key influence of the gas component on effects of
bars displacement and occurrence of asymmetry in isolated disk
structure in the whole. The initial distribution of dispersion of
stars velocities and the parameters of spherical subsystem suppose
the slow formation of bar without the account of gas component.
However the amplitude of one-arm harmonic  ($m=1$) is very small
and formation of the lopsided-disk does not occur.

\begin{figure}[!t]
\centerline{\includegraphics[width=1.0\hsize]{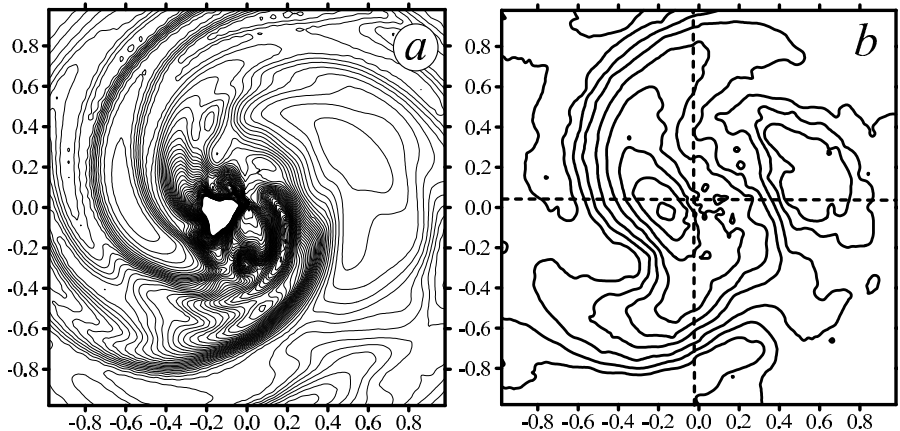}}
  \caption {
The isolines of surface density of gas (\textit{a}) and stellar
(\textit{b}) components are shown in the case $M_h/M_d=1$,
$M_{gas}/M_{d}=0.47$.
 The initial Toomre parameter is equal $Q_T\gee 1.3$ for the
 stellar disk in region $r\le 2L$.
 }\label{AVH-SG-Lops}
\end{figure}

The formation of harmonic  $m=1$ is possible in a massive cold gas
subsystem. The nonlinear interaction of a one-arm mode and
bar-mode ($m=2$) in the gas disk is the reason of asymmetry of
stellar disk also. The results of simulation in case of
 $M_{gas}=0.47 \cdot M_d$ in limits of $r\le 4L$ are shown
 in fig.~\ref{AVH-SG-Lops}.
With growth of relative gas mass we have amplifications of the bar
displacement concerning centre of disk and power of spiral
structure asymmetry. The considered mechanism of formation of
lopsided-galaxies is most effective in case of small halo mass and
if the halo scale exceeds the exponential disk scale in 2 times
and more.

\subsection{The problem of double-bars formation}

\begin{figure}[!t]
\centerline{\includegraphics[width=1.03\hsize]{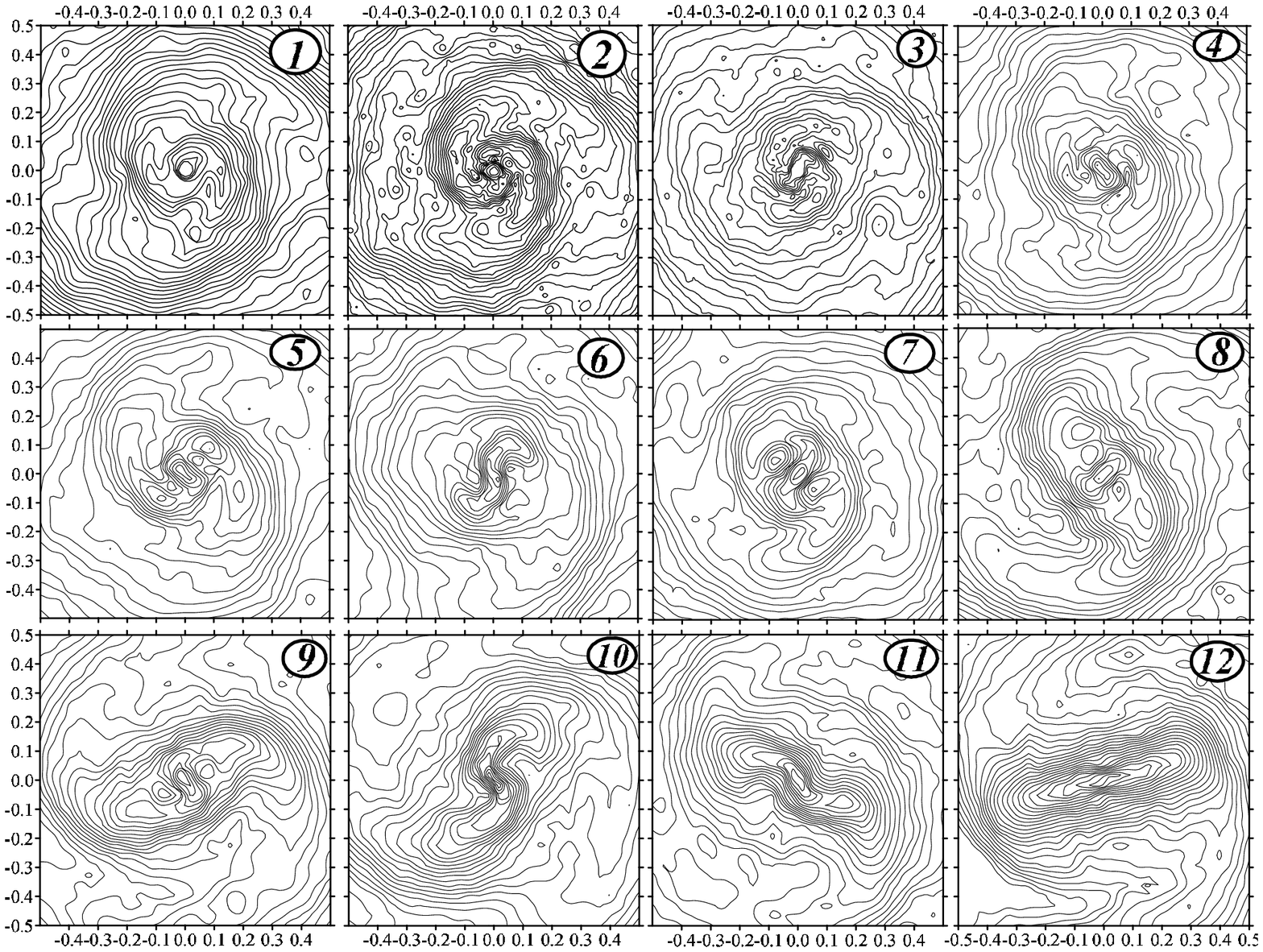}}
  \caption {
  The surface density distribution in different time (\textit{4}, \textit{5}, \textit{7}, \textit{8}, \textit{9} --- double bars).
 }\label{AVH-DB-dens}
\end{figure}

The bar formation requires not a hot initial disk and the halo
mass $M_{halo}+M_{bulge} \lee 2M_{disk}$ at halo scale a
$(1-4)\cdot L$ ($L$ --- exponential disk scale). The birth of
inner bar in numerical models occurs at presence of enough massive
bulge ($M_{bulge}\gee  0.3 M_{d}$). In Fig.\ref{AVH-DB-dens}, we
show the dis\-tri\-bu\-ti\-ons of surface density logarithm
$lg(\sigma)$ at the different time moments in model with
$M_{h}=M_{d}$, $M_{bulge}=0.6M_{d}$, in which at particular stages
there are structures such as double bars.

\begin{figure}[!t]
\centerline{\includegraphics[width=1.03\hsize]{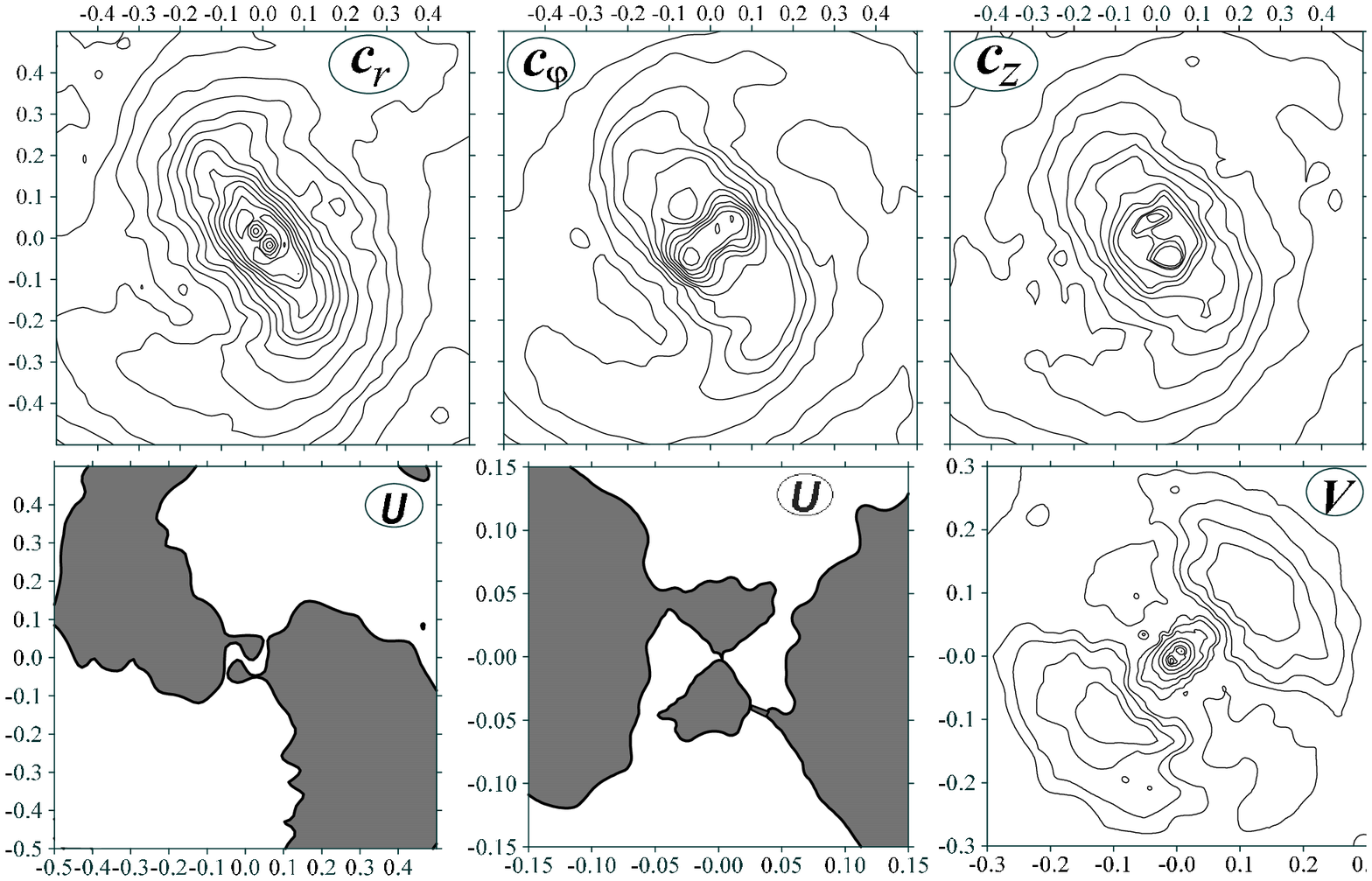}}
  \caption {
The distributions of the dispersions of velocity components
($c_r$, $c_\varphi$, $c_z$) and the isolines of azimuthal velocity
$V$ and radial velocity $U$. The areas of positive and negative
radial velocity are shown by different shading for the main and
inner parts of the stellar disk.
 }\label{AVH-DB-disp}
\end{figure}

The features of a kinematics of disk central region ($r <
2L$) at a stage of double bar are shown in a
fig.\ref{AVH-DB-disp}.
 The field of velocities in the stellar disk is the important
 information on existence of inner bar.
 The radial velocity $U$ demonstrates four-areal structure, both
 for inner bar, and for primary bar.

The lifetime of double-bar does not exceed 1--2 rotation periods
in the most ideal model of the stellar disk. The bending
instability of disk and/or bar is the important factor of
double-bars decay. The additional account of gas qualitatively
changes result. The double-bars are not forming in self-consistent
numerical stellar-gas models because of additional nonlinear
perturbations. External asymmetrical potential (the tidal
influence from the massive companion) gives similar result and
such models do not give double-bars also. The conclusion about a
dynamically fast phase of existence of ``double bar'' is agreed
with work [\ref{Moiseev-2002}], that the secondary bar not is real
dynamically allocated structure at observed galaxies, and
represent a combination of objects with various morphology.

\section{Conclusions}

\noindent 1. The bar formation in the hot gravitationally stable
stellar disk can be generated by the unstable cold gas disk. This
mechanism generates the bar even in case of the massive halo
$M_h/M_d\simeq 1-2$.

\noindent 2. The account of gas component strengthens the
formation of asymmetrical structures (lopsided, mode $m=1$) in a
isolated disk as a result of gravitational instability in case of
halo with small mass and large scale in comparison with a disk
scale. At Magellanic type galaxies the relation of halo mass to
disk mass in limits of optical radius on the average is less, than
at systems of early types.

\noindent 3. The self-consistent models with the double bar are
extremely unstable in relation to the various factors (transient
spiral waves in a disk plane, bar warps, bending instabilities of
a disk, tidal influence, gas component) at initial stages of
evolution. The conclusion about a very short phase of existence of
systems such as ``double bars'' (transient nature) is made and
similar structures can arise under special conditions at an
initial stage of bar-mode development.

 \medskip
 This work was supported by the Russian Foundation for Basic
Research through the grants RFBR~09-02-97021 and Technology
Program ``Research and Development in Priority Fields of Science
and Technology'' (contract 40.022. 1.1.1101).

\begin{center}
  \textbf{ References }
\end{center}

\begin{enumerate}

 \item\noindent\label{Bisnovatyi-Kogan-1984}
 Bisnovatyi-Kogan, G.S. 1984, Astrofizika, 20, 547

 \item\noindent\label{Friedli-1993}
Friedli D., Martinet L. Astron. Astrophys., 1993, 277, 27

 \item\noindent\label{Khoperskov-Moiseev-2001}
Khoperskov, A.V., Moiseev, A.V., Chulanova, E.A. 2001, Bull. SAO,
52, 135

 \item\noindent\label{Khoperskov-etal-2001}
 Khoperskov, A., Zasov, A., Tyurina, N. 2001, A.Rep, 45,
 180.

 \item\noindent\label{Khoperskov-2002}
 Khoperskov, A. 2002, Astr. Lett., 28, 651.

 \item\noindent\label{Moiseev-2001}
Moiseev, A.V. 2001, Bull. SAO,  51, 140

 \item\noindent\label{Moiseev-2002}
Moiseev, A.V. AstL, 2002, 28, 755

 \item\noindent\label{Odewahn-1991}
Odewahn, S.C. 1991, AJ, 101, 829

 \item\noindent\label{Pisano-2000}
Pisano D.J., Wilcots E.M., Elmegreen B.G. 2000, AJ, 120, 763

 \item\noindent\label{Polyachenko-1979}
 Polyachenko, V.L., Shukhman, I.G. 1979, SvA, 23, 407

 \item\noindent\label{Shlosman-1989}
Shlosman, I., Frank, J.,  Begelman, M.C. 1989, Nature, 338, 45

 \item\noindent\label{Zasov-2002}
Zasov, A.V., Khoperskov, A.V. 2002, Astron. Reports, 46,
 173

 \item\noindent\label{Vaucouleurs-1972}
 Vaucouleurs, de G., Freeman, K. 1972, Vistas Astron., 14, 163

\end{enumerate}

\end{document}